\documentclass[10pt,letterpaper]{article}
\usepackage{opex3}
\usepackage{amsmath}
\usepackage{amsfonts}
\usepackage{amssymb}
\usepackage[utf8]{inputenc}
\usepackage{dcolumn}

\newcolumntype{d}[1]{D{.}{.}{#1}}

\begin{document}

\title{Fast and optimal broad-band Stokes/Mueller polarimeter design by the use of a genetic algorithm}
\author{Paul Anton Letnes, Ingar Stian Nerbø, Lars Martin Sandvik Aas, Pål Gunnar Ellingsen, Morten Kildemo}
\address{Department of Physics, The Norwegian University of Science and Technology (NTNU), N-7491 Trondheim, Norway}
\email{paul.anton.letnes@gmail.com} 

\begin{abstract}
A fast multichannel Stokes/Mueller polarimeter with no mechanically moving parts has been designed to have close to optimal performance from $430-2000$ nm by applying a genetic algorithm. Stokes (Mueller) polarimeters are characterized by their ability to analyze the full Stokes (Mueller) vector (matrix) of the incident light (sample). This ability is characterized by the condition number, $\kappa$, which directly influences the measurement noise in polarimetric measurements. Due to the spectral dependence of the retardance in birefringent materials, it is not trivial to design a polarimeter using dispersive components. We present here both a method to do this optimization using a genetic algorithm, as well as simulation results. Our results include fast, broad-band polarimeter designs for spectrographic use, based on $2$ and $3$ Ferroelectric Liquid Crystals, whose material properties are taken from measured values. The results promise to reduce the measurement noise significantly over previous designs, up to a factor of 4.5 for a Mueller polarimeter, in addition to extending the spectral range. 
\end{abstract}


\section{Introduction}
\label{sec:Introduction}

Polarimeters are applied in a wide range of fields, from astronomy~\cite{gandorfer:1402,Collins2008,Alvarez-Herrero2010}, remote sensing~\cite{howe:202} and medical diagnostics~\cite{smith:210,Weinreb:95} to applications in ellipsometry such as characterizing gratings~\cite{Foldyna2009}, nanostructures~\cite{Nerbo:10} and rough surfaces~\cite{Jin2010,Germer2001,Germer2000}. As all polarimeters are based on inverting so-called system matrices, it is well known that the measurement error from independent Gaussian noise is minimized when the condition number ($\kappa$) of these system matrices is minimized~\cite{Stabo-Eeg2008a,Tyo2000}. It has been shown that $\kappa=\sqrt{3}$ is the best condition number that can be achieved for such a system, and that this optimal condition number can be achieved by several different approaches using various optical components (\emph{e.g.} rotating retarders~\cite{Sabatke2000}, division of amplitude~\cite{Azzam:03,Azzam1978}, and liquid-crystal variable retarders~\cite{Bueno2000}). In many applications it is necessary to perform fast spectroscopic measurements (\emph{e.g.} by using a Charge-Coupled Device (CCD) based spectrograph)~\cite{GarciaCaurel2004}. In that case, the wavelength dependence of the optical elements will cause the polarimeter not to be optimally conditioned over the full range simultaneously. A system based on two Ferroelectric Liquid Crystals (FLC) has been reported to be fast and reasonably well conditioned over the visible or near infrared spectral range~\cite{GarciaCaurel2004,Ladstein2007,Aas2010}. By introducing a third FLC a similar system has been proposed to have an acceptable condition number from the visible to the near infra-red ($430-1700$ nm)~\cite{Patent3FLC}. The design of a system having the best possible condition number over a broad spectrum is a challenging optimization problem due to the large number of parameters; many optimization algorithms are prone to return local optimums, and a direct search is too time consuming. To avoid this time-consuming exhaustive search, we decided to employ the Genetic Algorithm (GA). A GA simulates evolution on a population of individuals in order to find an optimal solution to the problem at hand. Genetic Algorithms were pioneered by Holland \cite{Holland:1992fk}, and are discussed in detail in \textit{e.g.} Ref.~\cite{Floreano2008}. GAs have previously been applied in ellipsometry to solve the inversion problem for the thickness and dielectric function of multiple thin layers, see \emph{e.g.} Ref.~\cite{Kudla2004804,Cormier:00,Fernandes:10}.

\section{Overdetermined polarimetry}
\label{sec:Overdetermined ellipsometry}

\begin{figure}
    \centering
    \includegraphics{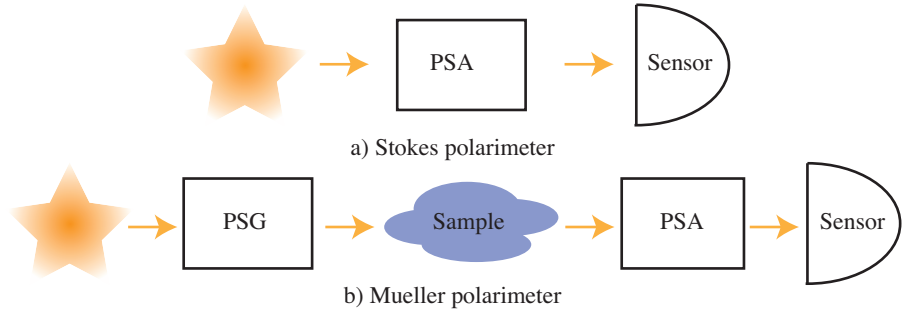}
    \caption{\label{fig:polarimeter-sketch} a) A Stokes polarimeter measures the polarization state of an arbitrary light source using a Polarization State Analyzer (PSA). b) A Mueller polarimeter measures how the polarization state of light, generated by with a Polarization State Generator (PSG), is changed by a sample.}
\end{figure}

A Stokes polarimeter consists of a polarization state analyzer (PSA) capable of measuring the Stokes vector of a polarization state, see Fig.~\ref{fig:polarimeter-sketch}. The PSA is based on performing at least 4 different measurements along different projection states. A measured Stokes vector $\mathbf{S}$ can then be expressed as $\mathbf{S}=\mathbf{A}^{-1}\mathbf{b}$, where $\mathbf{A}$ is a system matrix describing the PSA and $\mathbf{b}$ is a vector containing the intensity measurements. $\mathbf{A}^{-1}$ denotes the matrix inverse of $\mathbf{A}$, which in the case of overdetermined polarimetry with more than 4 projection states will denote the Moore--Penrose \emph{pseudoinverse}. The analyzing matrix $\mathbf{A}$ is constructed from the first rows of the Mueller matrices of the PSA for the different states. The noise in the measurements of $\mathbf{b}$ will be amplified by the condition number of $\mathbf{A}$, $\kappa_{\mathbf{A}}$, in the inversion to find $\mathbf{S}$. Therefore $\kappa_\mathbf{A}$ should be as small as possible, which correspond to do as independent measurements as possible (\emph{i.e.} to use projection states that are as orthogonal as possible). 

A Mueller matrix $\mathbf{M}$ describes how an interaction changes the polarization state of light, by transforming an incoming Stokes vector $\mathbf{S}_\text{in}$ to the outgoing Stokes vector $\mathbf{S}_\text{out}=\mathbf{M}\mathbf{S}_\text{in}$. To measure the Mueller matrix of a sample it is necessary to generate at least 4 different polarization states by a polarization state generator (PSG) and measure the outgoing Stokes vector by at least 4 measurements for each generated state. The measured intensities can then be arranged in a matrix $\mathbf{B}=\mathbf{AMW}$, where the system matrix $\mathbf{W}$ of the PSG contains the generated Stokes vectors as its columns. These generated Stokes vectors are found simply as the first column of the Mueller matrix of the PSG in the respective states. $\mathbf{M}$ can then be found by inversion as $\mathbf{M}=\mathbf{A}^{-1}\mathbf{BW}^{-1}$. The error $\Delta \mathbf{M}$ in $\mathbf{M}$ is then bounded by the condition numbers according to~\cite{Frantz1}

\begin{align}
\frac{\|\Delta \mathbf{M}\|}{\|\mathbf{M}\|}\lesssim \kappa_\mathbf{W}\kappa_\mathbf{A}\frac{\|\Delta \mathbf{B}\|}{\|\mathbf{B}\|}+ \kappa_\mathbf{A}\frac{\|\Delta \mathbf{A}\|}{\|\mathbf{A}\|}+\kappa_\mathbf{W}\frac{\|\Delta \mathbf{W}\|}{\|\mathbf{W}\|}.
\label{eq:error-propagation}
\end{align}
The condition number is given as $\kappa_\mathbf{A}=\|\mathbf{A}\|\|\mathbf{A}^{-1}\|$, which for the the 2-norm can be calculated from the ratio of the largest to the smallest singular value \cite{NumRes}. $\Delta \mathbf{A}$ and $\Delta \mathbf{W}$ are calibration errors, which increase with $\kappa$ when calibration methods using matrix inversion are applied. The PSG can be constructed from the same optical elements as the PSA, placed in the reverse order, which would give $\kappa_\mathbf{A}=\kappa_\mathbf{W}\equiv\kappa$. As the error in Mueller matrix measurements is proportional to $\kappa^2$, it is very important to keep this value as low as possible.  

\begin{figure}
    \centering
    \includegraphics{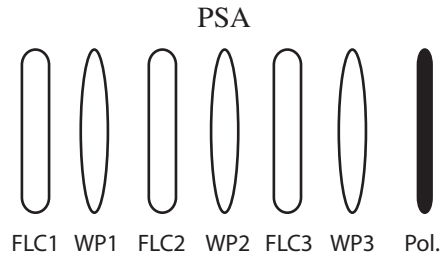}
    \caption{\label{fig:PSA-sketch} Sketch of a PSA consisting of 3 FLC's, 3 waveplates (WP), each with a retardance $\delta$ and an orientation $\theta$ relative to the transmission axis of a polarizer.}
\end{figure}

If 4 optimal states can be achieved (giving $\kappa=\sqrt3$), no advantage is found by doing a larger number of measurements with different states, compared to repeated measurements with the 4 optimal states \cite{Sabatke2000}. If, however, these optimal states can not be produced ($\kappa>\sqrt3$), the condition number, and hence the error, can be reduced by performing more than 4 measurements. For a FLC based polarimeter this can be done by using 3 FLCs followed by a polarizer as PSA, with up to 3 waveplates (WP) between the FLCs to increase the condition number (see Fig. \ref{fig:PSA-sketch}). A PSG can be constructed with the same elements in the reverse order. Since each FLC can be switched between two states (this switching can be described as a rotation of the fast axis of a retarder by $+45^\circ$), $2^3=8$ different states can be analyzed (generated) by the PSA (PSG). To accurately measure the Stokes vector, the system matrix $\mathbf{A}$ needs to be well known. For a Mueller polarimeter generating and analyzing 4 states in the PSG and PSA, the eigenvalue calibration method (ECM) \cite{Compain:99} can be applied. The ECM allows the measuring of the actual produced states by the PSA and PSG ($\mathbf{A}$ and $\mathbf{W}$), without relying on exact knowledge or modeling of the optical components. However, the ECM is based on the inversion of a product of measured intensity matrices $\mathbf{B}$ for measurements on a set of calibration samples. This product becomes singular for a system analyzing and generating more than four states. A workaround of this problem is to choose the subset of 4 out of 8 states which gives the lowest $\kappa$ value, and build a $\mathbf{B}$ matrix of those states to find 4 of the 8 rows (columns) of $\mathbf{A}$ ($\mathbf{W}$). More rows (columns) of $\mathbf{A}$ ($\mathbf{W}$) can then be found by calibrating on a different subset of the 8 states, giving the second lowest $\kappa$ value, and so on.
By repeating the calibration on different subsets of states, all the 8 rows (columns) of $\mathbf{A}$ ($\mathbf{W}$) can be found with low relative error $\|\Delta \mathbf{A}\| / \|\mathbf{A}\|$ $\left( \|\Delta \mathbf{W}\| / \|\mathbf{W}\| \right)$.

\section{Genetic optimization}
\label{sec:Genetic optimization}

\begin{figure}
    \centering
    \includegraphics{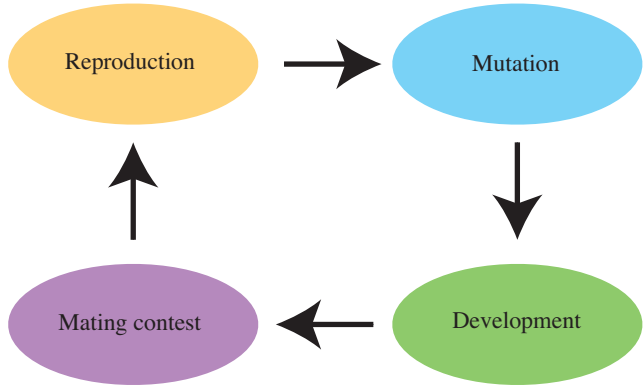}
    \caption{\label{fig:GA-processes} The four essential processes in a genetic algorithm are shown above. Sexual reproduction is performed by multi-point genetic crossover, giving rise to the next generation of individuals. Mutation can be simulated with simple bit negation (\emph{e.g.} $0 \rightarrow 1$ and \emph{vice versa}). Development is the process where a genotype is interpreted into its phenotype, \emph{i.e.} the binary genome is interpreted as a polarimeter design. In the mating contest, one evaluates the fitness of each individual's phenotype, and let the more fit individuals reproduce with higher probability than the less fit individuals.}
\end{figure}

In order to optimize $\kappa(\lambda)$, one can conceivably employ a variety of optimization algorithms, from simple brute-force exhaustive search to more advanced algorithms, such as \emph{e.g.} Levenberg--Marquardt, simulated annealing, and particle swarm optimization. Our group has previously performed optimization of a polarimeter design based on fixed components, namely, two FLCs and two waveplates. In this case, the optimization problem reduces to searching the space of $4$ orientation angles. With a resolution of $1^\circ$ per angle, this gives a search space consisting of $180^4 \approx 10^9$ states to evaluate; on modern computer hardware, this direct search can be performed. In order to optimize the retardances of the components as well, the total number of states increases to about $\left( 10^9 \right)^2 = 10^{18}$. Obviously, brute force exhaustive search is unfeasible for such large search spaces.

A GA performs optimization by simulating evolution in a population of individuals (here: simulated polarimeters). The three pillars of evolution are variation, heritability, and selection. Our initial population must have some initial genetic variation between the individuals; hence, we initialize our population by generating random individuals. Heritability means that the children have to carry on some of the traits of their parents. We simulate this by either cloning parents into children (asexual reproduction) or by performing genetic crossover (sexual reproduction) in a manner that leave children with some combination of the traits of their parents. Finally, selection is done by giving more fit individuals a larger probability of survival.\footnote{In the artificial intelligence literature, a large number of so-called selection protocols (algorithms) are discussed. This topic is too complex to be given a fair treatment here, but an introduction can be found in \emph{e.g.} Ref.~\cite{Floreano2008}. We will simply note that one must give also the lesser fit individuals a chance of reproduction, in order to maintain some variation and avoid premature convergence.} For a sketch of the essential processes involved in a GA, see Fig.~\ref{fig:GA-processes}.

Our GA builds directly on the description given by Holland~\cite{Holland:1992fk}, using a binary genome as the genetic representation. In this representation, a string of $0$s and $1$s represent the genome of the individual. To simulate mutation in our genetic algorithm, we employ logical bit negation; \emph{i.e.} $0 \rightarrow 1$ or vice versa. Sexual reproduction is simulated by using multi-point crossover, \emph{i.e.} simply cutting and pasting two genomes together, as described by Holland~\cite{Holland:1992fk}.

The interpretation of the genome into a phenotype (development), in this case a polarimeter design, is done in a straightforward way. For each variable in the polarimeter's configuration, \emph{i.e.} for each orientation angle and each retardance, we select $m$ bits in the genome (typically, $m = 8$) and interpret this number as an integer in the range from $1$ to $2^m$. The integer is subsequently interpreted as a real number in a predefined range, \emph{e.g.}, $\theta \in [0^\circ, 180^\circ]$. In order to avoid excessively large jumps in the search space due to mutations, we chose to implement the interpretation of bits into integers by using the Gray code, also known as the reflected binary code. The most important parameter values in our GA are shown in Table~\ref{tab:ga-parameters}. Making good choices for each of these parameters is often essential in order to ensure good convergence.

After determining the phenotype, we must assign to each simulated polarimeter individual a fitness function (also known as the objective function). In order to do this, we first calculate $\kappa (\lambda)$. As discussed, $\kappa^{-1}(\lambda)$ maximally takes on the value $1/\sqrt{3}$. Hence, we define an error function, $e$, as
\begin{align}
    \label{eq:error-function}
    e = \frac{1}{N_\lambda} \sum_{n=1}^{N_\lambda}
        \left(
            \kappa^{-1}(\lambda_n) - 1/\sqrt{3}
        \right)^4.
\end{align}
In Eq.~\eqref{eq:error-function}, $\lambda_n = \lambda_\text{min} + (n - 1) \Delta\lambda$, with $n = 1, 2, \ldots, N_\lambda$ and $\Delta\lambda = 5$ nm. $\lambda_\text{min}$ and $N_\lambda$ are determined by the wavelength range we are interested in. The choice of taking the difference between $\kappa^{-1}(\lambda)$ and the optimal value to power $4$ is done in order to ``punish'' peaks in the condition number more severely. As GAs conventionally seek to maximize the fitness function, we define an individual's fitness as
\begin{align}
    \label{eq:fitness}
    f = \frac{1}{e}.
\end{align}
This definition is convenient because $f$ takes on real and positive values where higher values represents more optimal polarimeter designs.

\section{Results}
\label{sec:Results}
For the case of a polarimeter based on $3$ FLCs and $3$ WPs, we have minimized $\kappa(\lambda)$ by varying the orientation angle, $\theta$, and the retardance, $\delta$, of all the elements. This yields a $12$-dimensional search space, \emph{i.e.}, $6$ retardances and $6$ orientation angles. $\theta$ is the angle between the fast axis of the retarder (WP or FLC) and the transmission axis of the polarizer (see Fig.~\ref{fig:PSA-sketch}), taken to be in the range $\theta \in [0^{\circ}, 180^{\circ}]$.  The retardance, $\delta$, is modeled using a modified Sellmeier equation,
\begin{align}
    \delta \approx 2\pi L
        \left[
            \frac{A_{UV}}
                {(\lambda^2 - \lambda_{UV}^2)^{1/2}}
            - \frac{A_{IR}}
                {(\lambda_{IR}^2 - \lambda^2)^{1/2}}
        \right],
\end{align}
where $A_{UV}$, $A_{IR}$, $\lambda_{UV}$, and $\lambda_{IR}$ are experimentally determined parameters for an FLC ($\lambda/2 @ 510$ nm, Displaytech Inc.) and a Quartz zero order waveplate ($\lambda/4 @465$ nm) taken directly from Refs.~\cite{Ladstein2007} (for the FLCs, $A_{IR}=0$). $L$ is a normalized thickness, with $L=1$ corresponding to a retardance of $\lambda/2 @ 510$ nm for the FLCs and $\lambda/4  @465$ nm for the waveplates. Each $L$ and $\theta$ are represented by $8$ bits each in the genome. We use experimental values to ensure that our design is based on as realistic components as possible.

\begin{figure}[bp]
    \centering
    \includegraphics{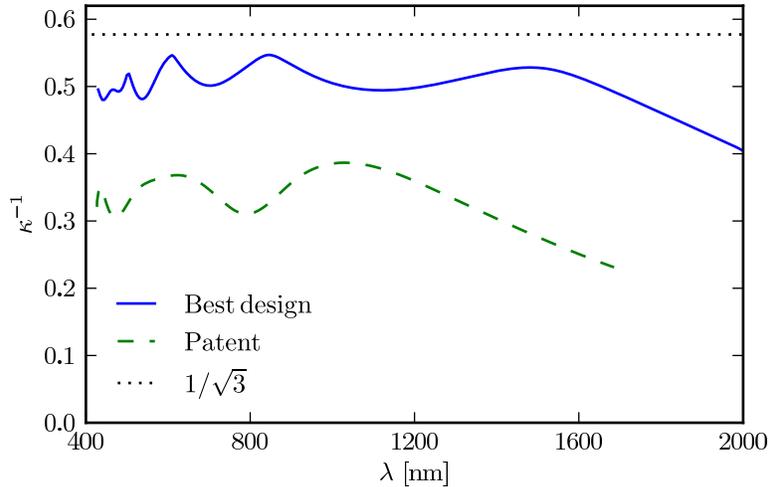}
    \caption{Inverse condition number for the best GA-generated $3$-FLC design. For comparison, we show the inverse condition number of the patented $3$-FLC design \cite{Patent3FLC}.}
    \label{fig:Condition-number}
\end{figure}

\begin{table}[btp]
    \centering
    \caption{\label{tab:optimal-3flc-polarimeter}Orientation angles, $\theta$, and normalized thicknesses $L$, of the components of the best $3$-FLC polarimeter. (WP = (fixed) waveplate)}
    \begin{tabular}{ld{-1}d{-1}}
    \hline
    Component   & \multicolumn{1}{c}{$\theta [^\circ]$} & \multicolumn{1}{c}{$L$} \\
    \hline
    FLC1    & 56.5  & 2.44 \\
    WP1     & 172.9 & 1.10 \\
    FLC2    & 143.3 & 1.20 \\
    WP2     & 127.1 & 1.66 \\
    FLC3    & 169.4 & 1.42 \\
    WP3     & 110.1 & 4.40 \\
    \hline
    \end{tabular}
\end{table}

The $3$-FLC polarimeter design scoring the highest fitness function is shown in Table~\ref{tab:optimal-3flc-polarimeter}. The wavelength range for which we optimized the polarimeter was from $430$ to $2000$ nm. To visualize the performance of this design, we show a plot of $\kappa^{-1}(\lambda)$ in Fig.~\ref{fig:Condition-number}. The inverse condition number, $\kappa^{-1}$, is larger than 0.5 over most parts of the spectrum, which is close to the optimal inverse condition number ($\kappa^{-1}=1/\sqrt{3}=0.577$). This is a great improvement compared to the earlier reported $3$-FLC design~\cite{Patent3FLC}, which oscillates around $\kappa^{-1}\approx 0.33$. The new design promise a decrease in noise amplification by up to a factor of $2.1$ for a Stokes polarimeter, and up to factor of $4.5$ for a Mueller polarimeter. In addition the upper spectral limit is extended from $1700$ nm to $2000$ nm. Shorter wavelengths than $430$ nm were not considered as the FLC material will be degraded by exposure to UV light. Previous designs often suffer from $\kappa^{-1}(\lambda)$ oscillating as a function of wavelength, whereas our solution is more uniform over the wavelength range we are interested in. This uniformity in $\kappa(\lambda)$ will, according to Eq.~\eqref{eq:error-propagation}, give a more uniform noise distribution over the spectrum.

\begin{figure}[btp]
    \centering
    \includegraphics{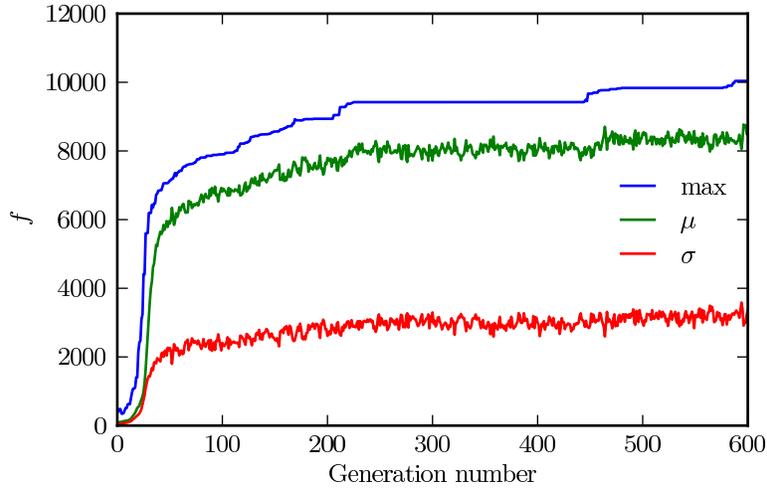}
    \caption{Convergence of fitness as a function of generation number. $\mu$ and $\sigma$ refer to the average and standard deviation of the population's fitness, respectively. The best result from this simulation is the one shown in Fig.~\ref{fig:Condition-number}.}
    \label{fig:GA-convergence}
\end{figure}

To give some idea of how fast the GA converges, a plot of $f$ (see Eq.~\eqref{eq:fitness}) as a function of the generation number is shown in Fig.~\ref{fig:GA-convergence}. The mean population fitness ($\mu$) and standard deviation ($\sigma$) is also shown. As so often happens with genetic algorithms, we see that the maximal and average fitness increases dramatically in the first few generations. Following this fast initial progress, evolution slows down considerably, before it finally converges after 600 generations. The parameters used in our GA to obtain these results are shown in Table~\ref{tab:ga-parameters}.

\begin{table}[tbp]
    \centering
    \caption{\label{tab:ga-parameters} Genetic Algorithm parameters. The ``crossover rate'' is the probability for two parents to undergo sexual reproduction (the alternative being asexual reproduction). The parameter ``crossover points'' refer to the number of points where we cut the genome during crossover (sexual reproduction). ``Mutation rate'' is the probability for any given individual to undergo one or several bit flip mutations in one generation.}
    \begin{tabular}{ld{-1}}
    \hline
    Parameter           & \multicolumn{1}{l}{Value} \\
    \hline
    Crossover rate      & 0.7 \\
    Crossover points    & 2 \\
    Mutation rate       & 0.2 \\
    Population size     & 500 \\
    \hline
    \end{tabular}
\end{table}

\begin{figure}[btp]
    \centering
    \includegraphics{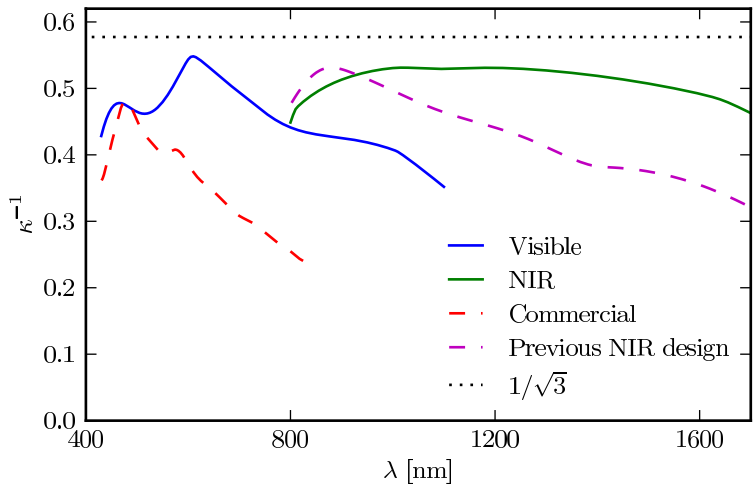}
    \caption{Condition number for two designs using $2$ FLC retarders and $2$ waveplates. By optimizing $\kappa(\lambda)$ over a narrower part of the spectrum, we can design good polarimeters with fewer components. The polarimeter designs labeled ``Visible'' and ``IR'' show our two designs, optimized for $430 \,\, \mathrm{nm} < \lambda < 1100 \,\, \mathrm{nm}$ and $800 \,\, \mathrm{nm} < \lambda < 1700 \,\, \mathrm{nm}$, respectively. For comparison with our ``NIR'' design, we show the previous simulated design from Ref.~\cite{Ladstein:2008uq}. The curve labeled ``Commercial'' shows the measured condition number of a commercial instrument (MM16, Horiba, 2006) based on the same (FLC) technology.}
    \label{fig:Condition-number-2-flc}
\end{figure}

A design using fewer components, in particular $2$ FLCs and $2$ waveplates, does have advantages. These advantages include increased transmission of light, as well as reduced cost and complexity with respect to building and maintaining the instrument. In addition some applications have weight and volume restrictions \cite{Alvarez-Herrero2010}. For these reasons, we have performed genetic optimization of the $2$-FLC design. In Fig.~\ref{fig:Condition-number-2-flc}, we show the performance of two polarimeter designs for the wavelength ranges $430-1100$ nm (compatible with an Si detector) and $800-1700$ nm. Both of these polarimeter designs show condition numbers which are considerably better than previously reported designs. The numerical parameters of the two designs based on $2$ FLCs are shown in Table~\ref{tab:optimal-2flc-polarimeter}.

\begin{table}[btp]
    \centering
    \caption{\label{tab:optimal-2flc-polarimeter}Orientation angle, $\theta$, and normalized thickness, $L$, of the $2$-FLC polarimeters shown in Fig.~\ref{fig:Condition-number-2-flc}.}
    \begin{tabular}{ld{-1}d{-1}d{-1}d{-1}}
    \hline
                & \multicolumn{2}{c}{\text{Visible design}} & \multicolumn{2}{c}{\text{NIR design}} \\
    Component   & \multicolumn{1}{c}{$\theta [^\circ]$} & \multicolumn{1}{c}{$L$} & \multicolumn{1}{c}{$\theta [^\circ]$} & \multicolumn{1}{c}{$L$} \\
    \hline
    FLC 1   & 90.4 & 1.17 & 177.9 & 2.60 \\
    WP 1    &  3.5 & 3.58 & 112.9 & 2.94 \\
    FLC 2   & 92.5 & 1.02 &  74.8 & 1.75 \\
    WP 2    & 19.8 & 3.52 & 163.1 & 4.71 \\
    \hline
    \end{tabular}
\end{table}

Our optimization algorithm can, with little effort, be applied to a wider range of polarimeter design. Any optical component can be included into our GA; for example, one can include fixed waveplates of different materials, prisms, mirrors, and other types of liquid crystal devices. The material of each component could also be a variable, which could help alleviate the dispersion problem. The only requirement is that the retardance of the component in question must be possible to either model theoretically or measure experimentally. It is possible to optimize a polarimeter for a different wavelength range, simply by changing program inputs. Focusing on a wavelength range which is as narrow as possible typically results in higher condition numbers than reported here. Evaluating different technologies, materials and components for polarimetry should thus be relatively straightforward. The task is not computationally formidable: we have used ordinary desktop computers in all our calculations.

\section{Conclusion}
\label{sec:Discussion and conclusion}

In conclusion, we have used genetic algorithms to optimize the design of a fast multichannel spectroscopic Stokes/Mueller polarimeter, using fast switching ferroelectric liquid crystals. We have presented three polarimeter designs which promise significant improvement with respect to previous work in terms of noise reduction and spectral range. Our approach requires relatively little computational effort. One can easily generate new designs if one should wish to use other components and materials, or if one wishes to focus on a different part of the optical spectrum. We hope that our designs will make polarimetry in general, and ellipsometry in particular, a less noisy and more efficient measurement technique.

\section*{Acknowledgements} 
\label{sec:Acknowledgements}
The authors would like to thank professor Keith Downing at the Department of Computer and Information Science at NTNU for helpful discussions regarding genetic algorithms and their implementation.

\end{document}